\documentclass{elsart3p}
\usepackage{amssymb}

\begin{document}

\begin{frontmatter}



\title{Effect of impurity pinning on conduction and specific heat in the Luttinger liquid}


\author{S.N. Artemenko, S.V. Remizov, D.S. Shapiro, R.R. Vakhitov}

\address{V.A. Kotelnikov Institute for Radio-engineering and Electronics of the RAS, %Russian Academy of Sciences, 
Mokhovaya str. 11-7, Moscow 125009, Russia}

\begin{abstract}
% Text of abstract
We study theoretically two effects related to impurity depinning that are common for 1D Luttinger liquid (LL) and linear-chain CDW conductors. First, we consider the electron transport through a single impurity in a 1D conducting channel and study a new regime of conduction related to LL sliding at voltage above a threshold one. The DC current in this regime is accompanied by oscillations with frequency $f =I/e$. This resembles the CDW depinning in linear-chain conductors, the Josephson effect, and the Coulomb blockade. Second, we found that strong pinning of the LL by impurities leads to a magnetic field dependence of the low-temperature specific heat similar to that observed experimentally in CDW compounds. We interpret this in favor of possibility of formation of the LL in linear-chain compounds.
\end{abstract}

\end{frontmatter}

\section{Introduction}
\label{Int}

In contrast to 3D electronic systems where electrons form, typically, a Fermi liquid, electrons in 1D systems are known to form a Luttinger liquid (LL) \cite{Giamarchi}. In case of the long-range Coulomb interaction the LL can be considered as a 1D version of the Wigner Crystal \cite{Schulz}. So the LL can be viewed as a simple prototype of an electronic crystal. A common feature of the LL with repulsive inter-electronic interaction and of other electronic crystals, like Wigner Crystal or charge-density wave (CDW), is their pinning by impurities. It is well-known that impurities in 1D systems with repulsive inter-electronic interaction form effectively large barriers and strongly suppress the current that leads to a power-law dependence of conductivity on voltage and/or on temperature~\cite{KaneFisher}. This effect can be described in terms of macroscopic tunnelling between different minima of a washboard potential describing interaction with an impurity  \cite{Giamarchi}. In Sec.~\ref{depinning} we study a new regime of conduction in 1D LL that corresponds to depinning of LL and its sliding at higher voltages corresponding to a large tilt of the washboard potential permitting depinning of the LL from the impurity. The DC current $\bar I$ in this regime is accompanied by coherent oscillations with frequency $f = \bar I/e$.  The effect is similar to the  well-known electric field depinning accompanied by generation of the narrow-band noise in linear-chain CDW compounds \cite{Monceau}.

The concept of the Luttinger liquid can be applied to quasi-1D CDW compounds as well. It was shown theoretically \cite {BrazFin,ArtNat} that formation of the CDW in quasi-1D materials with conducting chains coupled by elastic forces can be described in the frame of the LL if the interaction between the electrons is strong enough. And, in fact, the electronic interaction in these materials is, typically quite strong. There are also experimental indications to LL-like behavior of linear-chain compounds: a transition to a low temperature state characterized by LL-like conductivity was detected in focused-ion beam processed crystals of TaS$_3$ and NbSe$_3$ \cite{ZZPM}. In order to account for such behavior the possibility of stabilization of the LL state by defects in quasi-1D metals was suggested \cite{AR}. In general, whereas the main properties of quasi-1D conductors at higher temperatures are well understood, at low temperatures they demonstrate many intriguing properties which are not yet explained convincingly, e.g., unexpected strong magnetic field dependence of the specific heat in non-magnetic CDW materials TaS$_3$ and blue bronze \cite{Lasjaunias1,Staresinic,Lasjaunias2}. In Sec.~\ref{heat} we study the specific heat of pinned CDW described in terms of the LL, and find a magnetic field dependence of the heat capacity similar to that observed in the CDW materials. This dependence originates from strong pinning of the LL by impurities, and on partial violation of spin-charge separation by impurities resulting in possibility of  localization of an extra electron spin in segments between impurities. 

\section{Depinning of the Luttinger liquid from an impurity in 1D conducting channel}
\label{depinning}

The LL Hamiltonian describing interaction with impurities at $x=x_i$ reads~\cite{Giamarchi}
\begin{eqnarray}
&&
H_{imp} = -\frac{e}{\pi}\sum_i \int dx W \cos (\sqrt{2}\hat\Phi_\rho (t,x) -2k_Fx_i) \nonumber
\\
&&
\times \cos \sqrt{2}\hat\Phi_\sigma (t,x)\, \delta (x-x_i), 
\label{impurity}
\end{eqnarray}
where $\hat\Phi_\nu$ with $\nu=\rho, \sigma$ are the bosonic displacement fields in charge (spin) sectors, and $W$ is related to $2k_F$-component of the impurity potential.

In this section we study the response of the repulsive LL with single impurity at $x=0$ to DC voltage. For brevity we consider first the spinless LL using the Tomonaga-Luttinger (TL) model with short range interaction between electrons. Such interaction describes gated quantum wires where the long-range part of the interaction is screened by 3D gate electrodes. At the end we discuss essential modifications induced by the long-range Coulomb interaction and spin. 

Periodicity of the interaction potential is  associated with Friedel oscillations induced by impurity. Note that impurity pinning of the CDW is also related to Friedel oscillations~\cite{Zawadovsky}. The particle density operator in LL is expressed as $\hat \rho = -\frac{1}{\pi} \frac{\partial \hat\Phi_\rho}{\partial x} + \frac{k_F}{\pi }  \cos{ (2k_F x - 2 \hat\Phi_\rho)}$, where the first term describes smooth variations of the particle density. Fluctuations of $\hat\Phi$ make the expectation value of the rapidly oscillating second term in a free LL equal to zero. However, near the impurity the second term survives resulting in Friedel oscillations, {\it i. e.}, in $2k_F$ modulation of charge density around an impurity. As long as the current operator in the LL is given by relation $\hat I = (e /\pi)\partial_t \hat \Phi_\rho$  the current flow through the impurity means an increase of $\hat\Phi_\rho$ with time, which corresponds to a shift of the Friedel oscillations. So the Friedel oscillations is an obstacle for the current flow through an impurity. The well-known power-law I-V curves~\cite{Giamarchi,KaneFisher} are induced by tunneling between minima of washboard potential (\ref{impurity}) slightly inclined by an external voltage. This resembles Josephson junctions where fluctuations result in a finite voltage drop at currents below the Josephson critical current, $\bar I < I_c$ (for a review see Ref.~\cite{JosMQT}). However, in superconducting junctions at $\bar I > I_c$ the Josephson oscillations start, and this corresponds to an increase of the superconducting phase difference with time in the washboard potential inclined to an amount exceeding the critical value. Below we show that a similar regime must occur in the LL with an impurity when the applied voltage exceeds a threshold slope of the washboard potential at which the system can roll out from the minimum. Above the threshold the current is larger than the tunneling current in the sub-threshold regime corresponding to power-law I-V curves and is accompanied by oscillations with the washboard frequency $f = \bar I/e$.

%\section {Calculations}

The spinful Hamiltonian includes the impurity term (\ref{impurity}), the standard TL Hamiltonian of interacting electrons
\begin{equation}
H_{TL} =  \sum_{\nu=\rho, \sigma} \int dx \frac{\hbar \pi v_{F}}{2} \left[ 
\hat\Pi_\nu^2 +
\frac{1}{\pi^2 K_\nu^2} (\partial_x \hat\Phi_\nu)^2\right], \label{H0} 
\end{equation} 
and the term with an external electric field $E$ 
$$H_E= -\int dx \frac{\sqrt{2}e}{\pi} E
\hat\Phi_\rho,$$
where $K_\rho$  is  the LL parameter measuring the strength of the interaction: $K_\rho <1$ for repulsive and $K_\rho >1$ for attractive interaction. For a spin-independent electronic repulsion $K_\sigma =1$. 

Now we return to a spinless case and till the end of this section drop the subscript at $\hat\Phi_\rho$. Commuting field operators with the Hamiltonian we derive equation of motion for the Heisenberg operator $\hat\Phi$
\begin{eqnarray}
&&
\left(  v_\rho^2 \partial^2_{x} - \partial^2_{t} - \gamma \partial_{t}\right)\hat\Phi (t,x) = 
\label{phi}
\\
&&
\frac{e v_F}{\hbar}\left[2 W  \sin 2 \hat\Phi_0(t) \delta (x) - E \right],
\nonumber
\end{eqnarray}
where  $\hat\Phi_0(t) \equiv \hat\Phi(t, x=0)$, $v_\rho = v_F / K\rho$ is the plasmon velocity, and $\gamma$ describes small damping induced by coupling of electrons to a dissipative bosonic bath (to phonons or to density fluctuations in a metallic gate~\cite{Guinea,Natter}). 
Using Eq.~(\ref{phi}) with boundary conditions at the contacts one can express $\hat\Phi(t,x)$ in terms of its value at the impurity site. We apply boundary conditions~\cite{Grabert} derived for a wire adiabatically connected to ideal Fermi-liquid reservoirs at $x= \pm L/2$ with voltage difference $V$. We consider a long wire, $L \gg v_F/\gamma$, and neglect reflection of current pulses generated by the impurity from the contacts. Then we obtain equation of motion for $\hat\Phi_0(t)$ at the impurity position supplemented by relation between time averaged values
\begin{eqnarray}
&&
\partial_t \hat\Phi_0 (t) +  \int_0^\infty dt_1 e Z(t-t_1) \sin 2 \hat\Phi_0 (t_1) = \frac{\pi}{e} \bar I,  
\label{phi-t0}  
\\
&&
\bar I = G_0 (V-V_i), \; Z (t) = W_{i} K_\rho \int  \frac{d\omega }{2\pi}e^{- i\omega t} \sqrt{\frac{\omega}{\omega + i \gamma}} ,\nonumber
\end{eqnarray} 
where $\bar I = e^2 E v_F/\pi \hbar \gamma$ is the DC current, $G_0 = e^2/h$ is the conductance quantum per spin orientation, and 
\begin{equation}
V_i =2W \langle \sin 2 \hat\Phi_0(t) \rangle_t \label{I0} 
\end{equation}
is the DC voltage drop at the impurity.

As $\hat \Phi_0 (t)$ is an operator, it is not easy to solve Eq.~(\ref{phi-t0}). However, in the limit of strong inter-electron interaction ($K_\rho \to 0$) when fluctuations of $\hat \Phi_0 (t)$ at the impurity are small, $\hat \Phi_0 (t)$ can be treated as an ordinary function. So we consider first
the simple case $K_\rho \to 0$ when fluctuations can be neglected and Eq.~(\ref{phi-t0}) can be easily solved analytically. When $V \leq V_T = 2W$ the solution is stationary, $2\Phi_0 = \arcsin (U/V_T)$, with zero current $I =0$. 
At $V > V_T$ the solution is oscillatory with fundamental frequency  $f = \bar I /e $
\begin{equation}
\partial_t  \Phi_0 (t) \! = \! \frac{\pi \bar I^2/e}{ \sqrt{\bar I^2 \! + \!(G_0 V_T K_\rho )^2}\! +\! G_0 V_T K_\rho \sin \frac{2\pi \bar I}{e}  t}.
\label{dphi0}
\end{equation}
Eq.~(\ref{dphi0}) determines the current at the impurity site, $I (t, x=0) = e \dot \Phi_0 (t) / \pi $. Current at the clean part of the channel is equal to $I(t,x) = e \partial_t \Phi_0 (t-|x|/v_\rho)$ at $|x| \ll v_F/\gamma$, and $I(t,x) = \bar I$ at large distances from the impurity. 

For the DC voltage drop at the impurity we obtain
\begin{equation}
V_i = \left[ \sqrt{\bar I^2 + (G_0 V_T K_\rho)^2}- \bar I\right] /(G_0 K_\rho).
\label{s}
\end{equation}
Thus the oscillatory regime starts at $V > V_T$ and the amplitude of the oscillations decreases as voltage increases decaying at large voltages $V \gg V_T$.

Now we consider the case of finite values of $K_\rho$.
Mean square fluctuations $\langle \delta \hat\Phi^2 \rangle$ can be calculated from Keldysh Green's function $D^K = -i \langle \{ \delta \hat\Phi(t),\delta \hat\Phi(t')\}_+  \rangle$ at $t=t'$. This function can be expressed via retarded and advanced Green's functions
$$D^{R(A)} (t,t') = \pm i\theta [\pm(t-t')]  \langle [ \delta \hat\Phi(t),\delta \hat\Phi(t')]_-  \rangle$$ by relation $D^K (t,t')=D^R (t,t')f(t')-f(t)D^A (t,t')$, where Fourier transform of $f$ is related to the distribution function of bosonic excitations $N (\omega)$, $f(\omega) = 1 + 2N (\omega)$. In the equilibrium state $N (\omega)$ is the Planck distribution function. At low temperatures, smaller than all characteristic energies of the system, one can neglect contribution of thermally excited excitations, then $f(\omega) = {\rm sign} (\omega)$ and  $D^K$ can be expressed via the retarded and advanced functions. Note that acting in this way we neglect the effect of non-equilibrium distribution of bosonic excitations.

Now we derive equations of motion for the retarded and advanced Greens functions of fluctuations. This can be done in a standard way multiplying Eq.~\ref{phi} by $\hat\Phi (t',x')$ from the left and from the right, and combining them in order to obtain corresponding Green's function after averaging. Then using the Fourier transformation we express Greens functions at the impurity site and get rid of the coordinate dependence of Green's function. After that using the statistically averaged equation (\ref{phi-t0}) we subtract expectation values and use the self-consistent harmonic approximation
$$
\sin 2 \delta \hat\Phi \to  \langle \cos 2\delta \hat\Phi \rangle 2 \delta \hat\Phi = e^{-2 \langle \delta \hat\Phi^2 \rangle} 2 \delta \hat\Phi,
$$
and arrive, finally, to close equations of motion for $D^{R(A)}$. As we need this equations at frequencies larger than the small damping constant we neglect damping $\gamma$ and employ $Z (t) = W K_\rho \delta(t-t_1)$. Then the equation for $D^{R}$ reads
\begin{eqnarray}
&&
\partial_t D^R +  \frac{2e}{\hbar} W K_\rho C(t)   D^R = -\frac{\pi K_\rho}{2} \delta (t-t'), \nonumber \\
&&
 C(t) \equiv  \cos 2 \Phi (t_1) \langle \cos 2\delta \hat\Phi (t_1) \rangle.
\label{DR}
\end{eqnarray}
This equation and the similar equation for the advanced Green's function can be easily solved analytically. The solutions are
\begin{equation}
D^{R(A)} = - \frac{\pi K_\rho}{2} \theta [\pm(t-t')]e^{ \mp \frac{2e}{\hbar}W K_\rho \int^{t}_{t'} C(t_1) dt_1 }.
\label{DRA}
\end{equation}
This gives us for $\langle \delta \hat\Phi^2 \rangle = \frac{i}{2}D^{K}(t,t)$
\begin{equation}
\langle \delta \hat\Phi^2 \rangle = \frac{K_\rho}{2} \int^{\infty}_{0} \frac{dt_1}{t_1} e^{ - \frac{2e}{\hbar}W K_\rho \int^{t_1}_{0} C(t-t_2) dt_2 }  .
\label{DK}
\end{equation}
Eq.~(\ref{DK}) must be solved self-consistently with the statistically averaged equation (\ref{phi-t0}) in order to find $\langle \delta \hat\Phi^2 \rangle$ as function of $\cos 2 \Phi_0$. First, we calculate the threshold voltage. In the sub-threshold region $C$ does not depend on time. Then calculating integrals in (\ref{DK}) and using definition of $C$ (\ref{DR}) we obtain
\begin{equation}
\langle \delta \hat\Phi^2 \rangle =\frac{K_\rho}{2(1-K_\rho)} \ln \frac{\Lambda}{2K_\rho eW \cos 2 \Phi_0} ,
\label{DPhi}
\end{equation}
where $\Lambda \sim p_F v $ is a large cut-off energy. In accordance with our previous statement we found that fluctuations vanish at $K_\rho \to 0$.
Substituting (\ref{DPhi}) in (\ref{I0}) we obtain equation for $\Phi_0$ in the sub-threshold regime
\begin{equation}
2 W  \left( \frac{2K_\rho eW \cos 2 \Phi_0 }{\Lambda}\right)^{\frac{K_\rho}{1-K_\rho}} \sin 2 \Phi_0 =  V.
\label{phi0st}
\end{equation}
It has solution at $V < V_T$ with
\begin{equation}
V_T = 2W  \left( \frac{2K_\rho^{3/2} eW}{\Lambda}\right)^{\frac{K_\rho}{1-K_\rho}}\sqrt{1-K_\rho}.
\label{UT}
\end{equation}
We see from (\ref{UT}) that the finite threshold voltage exists provided $K_\rho < 1$, and this is in accordance with the condition that the impurity is a relevant perturbation in case of repulsive inter-electronic interaction.
Note, however, that fluctuations were taken into account in the self-consistent harmonic approximation that does not work well at $K_\rho \to 1$.

It is not simple to take into account fluctuations analytically in a general time-dependent case, but this can be done easily near the threshold value when the DC current is small, $\bar I \ll G_0 V_T K_\rho$. In this case we find (\ref{dphi0}) and (\ref{s}) again, but with different value of the threshold voltage $V_T$ given by (\ref{UT}).

So far we considered the spinless LL. In the spinful LL the spin-charge separation is violated at the impurity. This leads to modification of the results, in particular, of the threshold voltage. For a spin-independent electronic repulsion Eq.~(\ref{UT}) is substituted by
$$%\begin{equation}
V_T = W  \left( \frac{2 eW}{\Lambda}\right)^{\frac{1+K_\rho}{1-K_\rho}}K_\rho^{\frac{1+K_\rho}{3K_\rho}}\sqrt{2(1-K_\rho)}.
%\nonumber %label{UT}
$$%\end{equation}
Note also that electron transport current is accompanied by pulses of spin current generated at the impurity.

We considered also the case of the long-range Coulomb repulsion, and find that it does not result in important qualitative difference. One of the distinctions is the smaller role of fluctuations. An estimate for the threshold field in this case gives with the logarithmic accuracy
$$ V_T = 2 W \exp\left (-\frac{2}{a}\sqrt{\ln\frac{e}{\epsilon d W}}\right ),
$$
where $d$ is the wire diameter, $a^2 = \frac{4e^2}{\pi \hbar v_F \epsilon}$ is the dimensionless parameter measuring the strength of the Coulomb interaction, and $\epsilon$ is a background dielectric constant. 

In conclusion, we found that above the threshold voltage $V_T$ the current through an impurity generates coherent oscillations with the fundamental frequency $f = \bar I/e$. If in addition there is also an AC applied voltage with frequency $f_{0}$ then an analog of the resonance Shapiro steps observed in Josephson junctions will appear on the I-V curves. In our problem these are the steps of a constant voltage, the fundamental step being located at the current value $\bar I = e f_0$. Characteristic frequencies of the oscillations are determined by the strength of the impurity potential. In semiconducting quantum wires typical values of the impurity potential can be of the order of several meV, so depending on the strength of the electronic repulsion the frequency may fall into gigahertz or terahertz frequency region. Direct application of our results to real systems are limited by voltages smaller than distances to other electronic subbands. The results can be modified also due to different coupling of the 1D system to 3D environment.

\section{Magnetic field dependence of low-temperature specific heat of quasi-1D conductors}
\label{heat}

Now we take into account formation of the CDW due to electron-phonon  coupling.
We describe free phonons and electron-phonon coupling are described in a standard way, free phonons being presented by an elastic displacement field $\varphi$ with a harmonic action. We need to keep only slowly varying part of the $2k_F$ component of $\varphi$ leading to the CDW transition, therefore, in most of the cases the
weak dependence of the phonon frequency $\omega_{ph}(\textbf{q})$ on longitudinal component of the momentum, $q_\|$, can be
neglected. Then we bozonize the action (confer, e.g., \cite{VSch}) and integrate out phonon degrees
of freedom. This leaves the effective inter-electron interaction term
\begin{eqnarray}
&&
S_{\textrm{int}}= \left(\frac{g}{\pi\alpha
}\right)^2
\sum_{n, n_1}\,\int \limits%_{x,\tau,\tau_1}
dx d\tau d\tau_1D_{\varphi}(\tau-\tau_1,n-n_1)  \nonumber \\
&&
\times \cos[\sqrt{2}\Phi_\rho(\tau,x,n) -\sqrt{2}\Phi_\rho(\tau_1,x,n_1)]  \nonumber \\
&&
\times \cos \sqrt{2} \Phi_\sigma (\tau,x,n) \; \cos \sqrt{2} \Phi_\sigma (\tau_1,x,n_1)
\label{eph}
\end{eqnarray}
where $g$ is the electron-phonon coupling constant, $\alpha$ is a small cut-off distance, and $D_{\varphi}$ is the free phonon propagator. This equation differs from that of Ref.~ \cite{VSch} by summation over the chain numbers $n$.

With such action formation of the CDW was studied in \cite{ArtNat}. In ou problem we must add the impurities (\ref{impurity}). We concentrate on a simple case of very strong inter-electronic interaction ($K_\rho \ll 1$) resulting in strong pinning and small fluctuations of the fields $\Phi_\nu$ around classical trajectories corresponding to the minima of the total action. Fot the case of the strong pinning we must find classical trajectories satisfying the boundary conditions at the impurity positions. Boundary conditions that deliver the minimum of the impurity action are
$$%\begin{equation}
\sqrt{2}\Phi_\rho (t,x_i) -2k_Fx_i = n \pi,   \quad
  \sqrt{2}\Phi_\sigma (t,x_i) = m \pi
%\label{cond-imp}
$$%\end{equation}
where $(n+m)$ must be an even integer. The letter condition means that the number of electrons in segments between the impurities is an integer. 

Consider first, for illustration, the case without the CDW when the electron-phonon coupling is absent. In this case the system is effectively broken into independent segments with bounded LL. The energy spectrum of the segments consists of zero modes and excitations, the energy of the zero modes being dependent on number of extra electrons at a segment confined by impurities (confer \cite{AR}). The lowest energy of the zero-mode state corresponds to number of extra electrons $\delta n = 0$ or $\delta n = \pm 1$ depending on the values of the phases at the impurity. The states with one extra or one missing electron behave similar to magnetic impurity since they can have spins of both directions, and this leads to magnetic field dependence of the specific heat. Exited states are separated from the zero mode states by energies of the order of $v_\nu/lK_\rho$, where $v_\nu = v_F/K_\nu$, $\nu = \rho, \sigma$, and $l$ is a typical distance between the impurities, so the excited states can be ignored if the temperature is low enough.

A similar picture takes place in the presence of the CDW, the main difference being related to a coordinate dependence of $\Phi_\rho$ and $\Phi_\sigma$ between the impurities since the phase perturbations will be localized near the impurities. An exact solution for the zero mode phase distribution depends on details of the crystalline structure and phonon dispersion. Therefore, we concentrate on the qualitative features and consider a simple model resembling approaches used to describe interaction of the CDW with impurity in Refs.~\cite{Larkin,BrazLar} (the spin-density wave case was treated similarly in \cite{Melin}). In these papers phase variation around an impurity was described by static sine-Gordon equation. Our case is somewhat different because we need to solve coupled equations for two phases $\Phi_\rho$ and $\Phi_\sigma$. Note that the details of phase variations are not important for our problem, furthermore, we will find that the magnetic field dependence of the specific heat is rather universal and does not depend on these details and on energies of the zero-mode states. So we use an oversimplified model analogous to that of Refs.~\cite{Larkin,BrazLar}. Namely, having in mind small impurity density we consider isolated impurities and neglect in action (\ref{eph}) variations of the phases at the chains that does not contain the impurity.  Matching the zero mode solutions of the equations for the phase variations at the impurity we found that the smallest energy can be achieved by the zero-mode states with number of extra electrons localized near the impurity $\delta n = 0, \pm 1$. These states correspond to variations of the phases $\delta\Phi_\rho$ and $\delta\Phi_\sigma$ around the impurity equal to $(0,0)$ and $(\pm\pi/\sqrt{2},\pm\pi/\sqrt{2})$, respectively. The former state has energy $E_{\rho,0} = \frac{2M}{K_\rho}(1-\cos{\phi_i})$ in the charge sector and $E_{\sigma,0} =0$ in the spin sector, where $M$ is an energy of the order of the CDW gap, and $\phi_i = k_Fx_i$. The latter states describing charge $\pm e$ and spin $\pm 1$ localized near the impurity have energy $E_{\rho,1} = \frac{M}{K_\rho}(2-\cos{\phi_i} - \sin{\phi_i})$ in the charge sector and $E_{\sigma,1} \sim M \ll E_{\rho,1}$ in the spin sector. In the presence of magnetic field the energy of the states with $\pm 1$ extra spin must be shifted by $\pm \mu_B H$, where $\mu_B$ is of the order of the Bohr magneton of an electron. The lowest excited states in the charge sector are separated by the energy gaps of the order of a phonon frequency $\omega_\rho \sim s_\perp/d$ where $s_\perp$ is the sound velocity in the direction perpendicular to the chains, and $d$ is an inter-chain distance. In the spin sector the excited states are separated by the energy $\omega_\sigma \sim M$. So for temperatures $T \ll \omega_\rho, \omega_\sigma$ the contribution of the excited states to the partition function and, hence, to the specific heat can be neglected.

Now we can calculate the partition function of the system taking into account that impurities are considered as independent, and treating the phases $\phi_i$ at the impurity positions as a randomly distributed variable. Then the specific heat of the sample with many impurities is related to the partition function $Z_i$ of a single impurity as 
$$%\begin{equation}
C = n_{imp} \int \frac{d\phi_i}{2\pi}  c_i, \quad c_i = - T \frac{\partial^2 F_i}{\partial T^2},\quad F_i = -T \ln Z_i
%\label{C}
$$%\end{equation}
where $n_{imp}$ is the impurity density. 

After simple calculations we find the contribution of a single impurity to the specific heat
$$%\begin{equation}
c_i = \frac{4 e^{u}\left[ 8h^2 e^{u} + (u + h)^2 e^h + (u - h)^2 e^{-h} \right] }{\left(1+ 4 e^{u} \cosh h  \right)^2}
%\label{c}
$$%\end{equation}
with $u= (E_{\rho,0} - E_{\rho,1} - E_{\sigma,1})/T$ and $h = \mu_B H/T$.

We are interested in the magnetic field dependence of the specific heat at low temperatures, $T \ll M$. Then typical values of $u$ for different impurities are large, $|u| \gg 1$, and the main contribution to the magnetic field dependent part of the specific heat is given by values of $\phi_i$ for which $u > 0$. This is natural because only those impurities can contribute to a magnetic field dependence of the specific heat for which the states with a localized spin have the energies smaller, than the spinless states. Finally, we find for the magnetic field-dependent part of the specific heat
\begin{equation}
C_H = b n_{imp} \frac{h^2}{\cosh^2 h}
 \label{cH}
\end{equation}
where $b$ is the fraction of the impurities with $u > 0$. Note that the shape of $C_H(h)$ is independent on the exact solutions for zero-mode states. In the simple model considered here $b \approx 1/2$, while in general case the factor $b$ can be different.

Magnetic field dependence (\ref{cH}) qualitatively agrees with the experimental data \cite{Lasjaunias1,Staresinic,Lasjaunias2}. Hysteretic behavior observed in the experiments can be understood as well, because one can expect different metastable distributions of electrons in segments between impurities. Application of magnetic field and heating/cooling cycles may easily result in transitions between such states with close energies.

In conclusion, we found that strong pinning of the LL and partial violation of the spin-charge separation by impurities leads to possibility of localization of electron spins by impurities resulting in magnetic field dependence of the specific heat similar to that observed experimentally. We consider this agreement as an evidence in favor of formation of the LL state in CDW conductors at low temperatures. Similar magnetic field dependence is possible also in the LL with defects without the CDW. 

% \appendix

% \section{}
% \label{}

\section*{ACKNOWLEDGMENTS}

The work was supported
by Russian Foundation for Basic Research. A part of the research was performed in the frame of the
CNRS-RAS-RFBR Associated European Laboratory ``Physical properties
of coherent electronic states in condensed matter'' between Institut N\'eel (CRTBT)
and IRE RAS.

\end{document}